\documentclass[letterpaper]{article}
\usepackage[T1]{fontenc}
\usepackage{geometry}
\usepackage{setspace}

\usepackage[style = chem-acs, doi = true, articletitle=true]{biblatex}
\usepackage{graphicx}
\usepackage{tabularx}
\usepackage{float}
\newfloat{scheme}{htbp}{los}
\floatname{scheme}{Scheme}
\floatname{chart}{Chart}
\newfloat{graph}{htbp}{loh}

\usepackage{chemformula} 

\usepackage[version = 4]{mhchem} 
\usepackage{amssymb}

\usepackage[colorlinks=true,
            linkcolor=blue,
            citecolor=blue,
            urlcolor=blue,
            pdfborder={0 0 0}]{hyperref}

\usepackage{authblk}
\author[1]{Felix Riccius}
\author[1]{Karsten Reuter}
\author[1,*]{Hendrik H. Heenen}

\author[2,\dag]{Jutta Rogal}

\affil[1]{Fritz-Haber-Institut der Max-Planck-Gesellschaft, Faradayweg 4-6, D-14195 Berlin, Germany}
\affil[2]{Initiative for Computational Catalysis, Flatiron Institute, New York 10010, New York, United
States}

\title{Finite-Temperature Thermodynamics of Cu(100) Oxidation: Missing-Row Reconstruction, Defect States, and Order-Disorder Transition from Nested Sampling}

\date{*Email: heenen@fhi.mpg.de, \dag Email: jrogal@flatironinstitute.org}

\begin{document}
\maketitle

\begin{abstract}

Metal surfaces undergo structural, compositional, and morphological changes in response to their chemical environment. Tuning the surfaces' function and stability for a given application correspondingly necessitates an understanding of how this surface evolution couples to external conditions. Here, we demonstrate the feasibility of nested sampling simulations to obtain this coupling at first-principles predictive quality. By exploring the full configuration space, nested sampling estimates the partition function and gives direct access to desired thermodynamic ensemble averages at any temperature without prior knowledge. Computational feasibility is achieved through machine-learned interatomic potentials, an efficient GPU implementation of the sampling algorithm and bespoke sampling moves. Applied to the early oxidation of Cu(100), the approach successfully predicts the experimentally observed, complex $(2\sqrt{2}\times\sqrt{2})$R45$^\circ$-O missing-row reconstruction. The full access to the partition function enables a detailed characterization of the temperature-dependent surface evolution, mapping the emergence of defect states and the order-disorder transition of the reconstructed surface.

\end{abstract}

\section*{Keywords}

Nested Sampling, Surface Reconstruction, \textit{Ab initio} Thermodynamics, Oxidation, Copper(100), Defects

\section*{Abbreviations}

MR = Missing Row; MRR = Missing Row Reconstruction; MC = Monte Carlo; MLIP = Machine-learned interatomic potentials

\section{Introduction}

The composition and morphology of transition metal surfaces are key determinants of their function and stability in applications ranging from optoelectronics and sensors to corrosion protection and heterogeneous catalysis.\cite{tao2024surface, Diebold2010, Shaikhutdinov2012, Yu2016,Maurice2018} Under {\em operando} conditions, these surfaces are prone to dynamic restructuring in response to the chemical environment, temperature, and pressure, which can substantially alter performance but also offers a route to its tuning for target applications.\cite{polo2019surface, chen2025modeling} Understanding how these surfaces reconstruct is a long-standing challenge for both experiment and simulations, yet their evolution with external conditions remains even further out of reach. Although decades of experimental characterization efforts have successfully unveiled increasingly complex surface reconstructions, defects, and domain boundaries \cite{salmeron2020high, zhu2024vacancy, gattinoni2015atomistic}, current state-of-the-art {\em in situ} and {\em operando} techniques are typically only capable of yielding a few snapshots at limited resolution, leaving an incomplete picture of both the surface structure and its evolution under operating conditions.\cite{chee2023operando} Furthermore, structural probes may be biased toward crystalline domains~\cite{jones2018maximising} and cannot reliably distinguish equilibrated reconstructions from kinetically trapped metastable structures~\cite{Reuter2005Handbook}. Computational surface science offers a route to close this gap by combining density functional theory (DFT) calculations with \textit{ab initio} thermodynamics to predict surface phase diagrams and identify stable reconstructions.\cite{reuter2001composition, rogal2006ab, reuter2016ab, lee2021completing, lee2024rise} Powerful as this approach is, the high computational cost of DFT restricts the configurational space to a handful of idealized structures selected by human intuition and evaluated at zero temperature. Unknown or disordered reconstructions therefore escape predictions, and their evolution as a function of external conditions remains inaccessible.

Extensive configurational sampling, which is needed as a basis for predictive modeling, has become increasingly feasible with the timely emergence of machine-learned interatomic potentials (MLIPs) trained on DFT data, providing accurate energies and forces at a fraction of the computational cost.\cite{mou2023bridging} For surface science applications, several sampling techniques have recently been combined with MLIPs, including grand canonical Monte Carlo,\cite{xu2022atomistic, kanno2025grand} virtual surface site relaxation-Monte Carlo,\cite{du2023machine, du2025accelerating} and replica exchange methods.\cite{zhou2019determining, riccius2025out, kanno2025grand} These approaches greatly expand the sampled configurational space, but each remains confined to preselected thermodynamic conditions\,---\,e.g.\ a single temperature or a narrow range\,---\,and typically requires prior knowledge of the relevant phases and transition temperatures. Nested sampling\cite{skilling2004nested, skilling2006nested, ashton2022nested, partay2021nested} overcomes this restriction by estimating the partition function directly and sampling a configurational ensemble that can be reweighted to arbitrary temperatures. A single simulation yields the full temperature dependence of derived thermodynamic quantities and ensemble averages over structural observables, as well as the characterization of phase transitions (most directly through peaks in the heat capacity). Importantly, nested sampling does not require prior assumptions about the relevant phases, transition temperatures, or structural motifs, greatly reducing the dependence of simulation results on the respective inputs. Although nested sampling is well established for bulk systems,\cite{partay2021nested, ungert2023neural, unglert2025replica, baldock2016determining} its adaptation to surfaces has so far been limited to model systems such as Lennard-Jones surfaces.\cite{yang2024surface} Here, we extend nested sampling to realistic systems by combining it with an MLIP and explore the physics that drives a non-trivial surface reconstruction.

Copper and its oxide surface reconstructions represent a compelling application due to its structural complexity and its particular relevance in heterogeneous catalysis.\cite{gattinoni2015atomistic, li2023converting, ali2022activity, sun2018cuprous} Here, we focus on the Cu(100) surface, which upon oxidation forms a $(2\sqrt{2}\times\sqrt{2})\;R45^{\circ}$-O missing-row reconstruction (MRR), that has been observed both experimentally\cite{harrison2006adsorbate, liu2024modeling} and theoretically.\cite{jacobsen1990theory, lian2017calculations, kangas2012formation} While MRR formation has been tied to the relief of oxygen-induced surface stress, \cite{harrison2006adsorbate,lahtonen2008oxygen, li2022uneven} its evolution with temperature, which produces a rich defect chemistry of orthogonally oriented rows as demonstrated by recent experiments, \cite{liu2024modeling, shields2024stm} remains unresolved. To obtain a detailed thermodynamic picture of the Cu(100)-O MRR, we combine nested sampling with an accurate MACE potential\cite{batatia2022mace} trained on an extensive set of DFT data. Nested sampling, however, requires a substantial number of energy evaluations, resulting in a very high computational cost. To address this challenge, we introduce two key extensions that render the simulations tractable: the augmentation of the Monte Carlo (MC) moves in the nested sampling approach and an efficient GPU implementation of the sampling algorithm together with the MLIP evaluation. Our simulations recover the MRR and confirm its formation as an oxygen-induced phenomenon. Beyond the ground-state structure, nested sampling captures the temperature-driven surface evolution, revealing defect states as thermodynamically favored rather than kinetically trapped and the proliferation of defects as a driving force of an order-disorder transition. These findings emphasize the strength of nested sampling to predict not only stable surface structures but their full evolution with external conditions, the quantity that ultimately governs function.

\newpage

\section{Computational Approach}

\subsection{The Nested Sampling Algorithm}
\label{sec:methods}

Nested sampling, originally introduced by Skilling~\cite{skilling2004nested, skilling2006nested} as an algorithm for computing Bayesian evidence while producing posterior samples as a byproduct, has emerged over the last decade as a powerful tool within the materials science community~\cite{partay2010efficient,partay2021nested, ungert2023neural, unglert2025replica, baldock2016determining}. Besides sampling the configuration space, a nested sampling simulation provides direct access to the partition function and, consequently, to all thermodynamic properties and atomic observables that follow from it.

Nested sampling is initialized with a set of $K$  \textit{walkers} uniformly distributed in configuration space (the \textit{live set}). Each walker corresponds to a replica of the system with configuration $\mathbf{x}_i \in \mathbb{R}^{3N}$, initialized with random, ideal-gas-like positions to ensure a uniform distribution over the accessible phase space. Consequently, initialization does not require prior knowledge of the system under study.

During the simulation, the phase space is explored by systematically reducing its accessible volume through an iterative procedure. First, the walker $\mathbf{x}_i$ with the highest energy in iteration $i$ is identified and used to set a new energy ceiling $E_{\mathrm{limit}} = E(\mathbf{x}_i)$. The phase space volume $\Gamma_i$ enclosing all states with $E < E_{\mathrm{limit}}$ is given by $\Gamma_i = \Gamma_0 \left[ K/(K+1) \right]^i$, where $\Gamma_0$ denotes the total phase space volume. Second, the configuration $\mathbf{x}_i$ is removed from the live set and added to the ensemble of nested sampling configurations. A new configuration is introduced to the live set by cloning one of the remaining walkers. To restore a uniform distribution, the cloned configuration is decorrelated through a series of MC steps subject to the energy constraint $E < E_{\mathrm{limit}}$.\cite{partay2021nested}

This two-step iterative procedure yields a sequence of sampled configurations $\mathbf{x}_i$ with energies $E(\mathbf{x}_i)$ and associated phase space weights $w_i = \Gamma_i - \Gamma_{i+1}$. This ensemble can be used to approximate the canonical partition function,
\begin{equation}
    Z_{N,V}(\beta) \approx \sum_i w_i \exp\!\left( -\beta E(\mathbf{x}_i) \right) \;, \label{eqn:partition}
\end{equation}
where $\beta = 1/(k_{\mathrm{B}}T)$ is the inverse temperature. From the partition function, thermodynamic properties, such as the heat capacity, are readily obtained,  and ensemble averages of any atomic observable are directly accessible for any temperature (see SI section 5.1).

The quality of the partition function estimate, and by extension of all derived quantities, depends strongly on the number of walkers $K$. Increasing $K$ yields a finer resolution of the phase space and more samples throughout the simulation, thereby improving the accuracy of the approximation. Funnel-like unimodal energy landscapes can be sampled adequately with relatively few walkers, however, more complex multimodal landscapes, which we expect for a surface system, require larger $K$ to cover all competing minima basins. Similarly, a larger particle count increases the dimension of the configuration space and usually also coincides with increased complexity and modality of the energy landscape, requiring larger $K$. An increase in walkers, however, entails a near linear increase in computational effort.~\cite{partay2021nested}

\subsection{Walker Decorrelation}

\begin{figure}[H]
    \centering
    \includegraphics[width=0.5\linewidth]{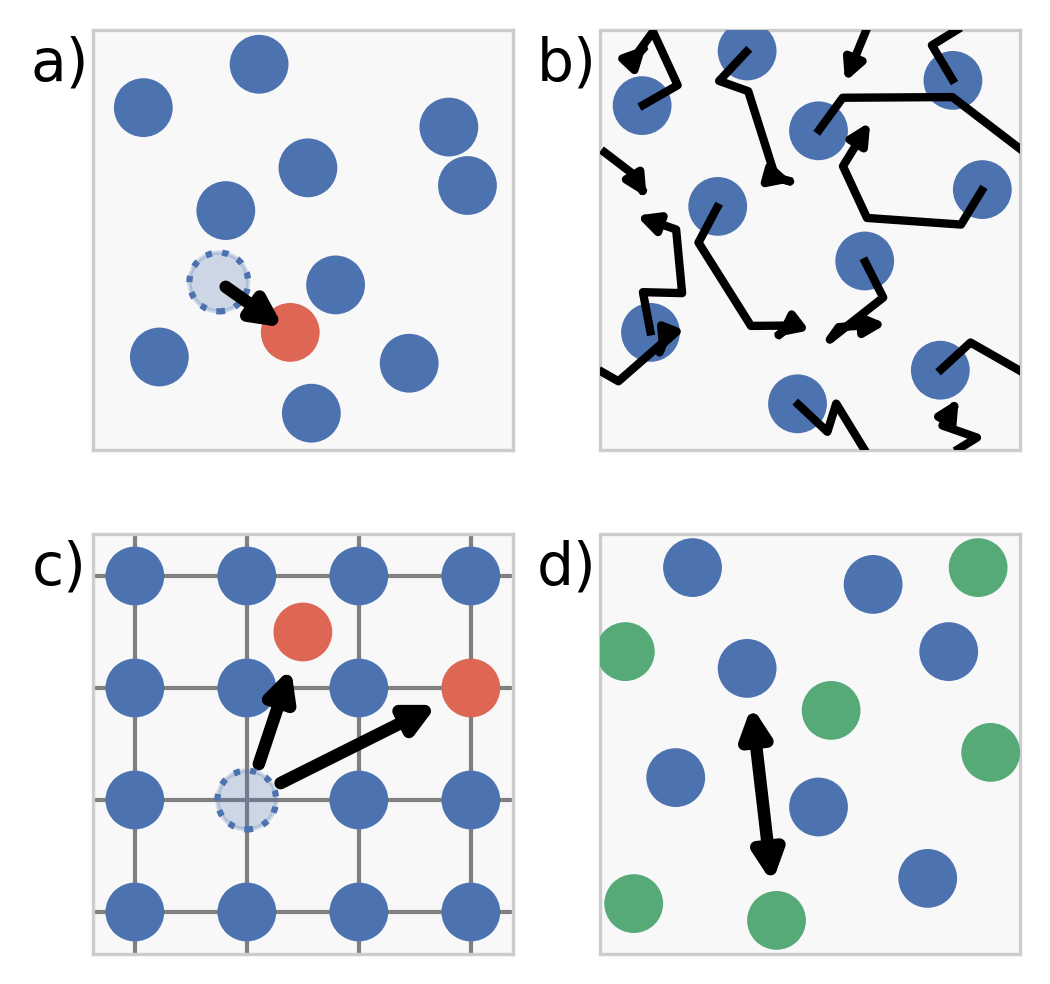}
    \caption{Illustration of the different MC move types employed for walker decorrelation: (a)~single-atom displacement, (b)~Galilean Monte Carlo (GMC), a collective move acting on multiple atoms, (c)~lattice-based single-atom displacement moves, and (d)~identity swap between different atomic species.}
    \label{fig:main_1}
\end{figure}

A crucial step in nested sampling simulations is the decorrelation of walkers to ensure uniform sampling of configurations within the confined phase space. Decorrelation is achieved through a multistep MC walk, in which the energy constraint $E < E_{\mathrm{limit}}$ serves as a hard acceptance criterion for each proposed move. For the studied Cu(100) system, the standard MC moves proved insufficient to effectively decorrelate walkers, necessitating the development of additional proposal moves. Overall, we employ four different MC move types, illustrated in Fig.~\ref{fig:main_1}: single-atom displacements (\ref{fig:main_1}a), Galilean Monte Carlo (GMC)~\cite{skilling2012bayesian} (\ref{fig:main_1}b), lattice-based MC moves (\ref{fig:main_1}c), and identity swaps (\ref{fig:main_1}d).

Single-atom displacements (Fig.~\ref{fig:main_1}a) are valid in principle but become increasingly inefficient as the number of atoms grows. Collective moves, which update all atomic coordinates simultaneously, offer a more practical alternative. In particular, GMC~\cite{skilling2012bayesian} (Fig.~\ref{fig:main_1}b), a multi-step collective move developed specifically for nested sampling, has gained popularity~\cite{ungert2023neural, yang2024surface} (see SI section 4 for details of the algorithm). However, the slab model of the Cu(100) surface with its fixed substrate layer (Fig.~\ref{fig:main_2}) contains well-defined adsorption sites that are separated by sizeable energy barriers. Because updates such as GMC use only small, continuous atom displacements, they effectively fail to cross these barriers once the maximum energy $E_{\rm limit}$ of the nested sampling simulation falls below the barrier energy. At this point, the phase space accessible to the simulation is partitioned into disconnected regions, where atoms on the surface are likely to get trapped in a given adsorption site. Consequently, the walkers become confined to subspaces, preventing their decorrelation. To alleviate this issue, we augment the MC walk with lattice-based atom displacement moves that exploit the crystallographic nature of the surface (see SI section 4). As illustrated in Fig.~\ref{fig:main_1}c, two different lattice moves are introduced: ({\it i})~displacement of an atom by $k$ lattice positions along the $x$- and $y$-directions, enabling direct transitions between neighboring adsorption sites; and ({\it ii})~displacement of an atom by $k{+}0.5$ lattice positions along the $x$- and $y$-directions combined with an up or down shift of one monolayer (ML) perpendicular to the surface, designed to overcome step edges on the surface. In addition to the lattice, GMC, and single-atom displacement moves, identity swaps between atomic species (that maintain the canonical ensemble) are included for the CuO simulations (see Fig.~\ref{fig:main_1}d). A more detailed description of the walker decorrelation procedure, including the move frequencies, can be found in the SI section 4.

Walker decorrelation is by far the most computationally demanding task of a nested sampling simulation. To cope with the high computational cost, particularly when combined with a GPU-native MLIP such as MACE, we implement the entire MC walk in \texttt{PyTorch}, following the \texttt{torch-sim} framework~\cite{cohen2025torchsim}, which includes the GPU-based \texttt{batch\_nl} neighbor list. Consequently, the complete MC walk is executed exclusively on the GPU, eliminating the overhead associated with repeated GPU--CPU data transfer. This implementation allows batching and parallelizing multiple MC walks on a single GPU, significantly increasing GPU utilization with minimal performance degradation from inter-walk interference which leads to a net $\approx 6\times$ speed up, making the required simulations computationally tractable.

\subsection{Simulation Setup}
The MRR on Cu(100) can be captured using comparatively small surface models constructed from multiples of the c$(2\times 1)$ unit cell. The energetics of the Cu(100) surface and its oxide reconstructions are described using a MACE machine-learned interatomic potential~\cite{batatia2022mace} trained on DFT reference data, following previous work on Cu surfaces~\cite{lee2021completing, riccius2025out}. Rather than training from scratch, we fine-tune a more general CuO MACE potential~\cite{riccius2025out} for our system using an active-learning scheme built around a genetic algorithm for data generation~\cite{lombardi2026ezga}. The resulting potential achieves a root-mean-square error (RMSE) of $1.7\,\mathrm{meV/atom}$ on energies and $45\,\mathrm{meV/\text{\AA}}$ on forces. More information on the training procedure is provided in SI section 1.

\begin{figure}[H]
    \centering
    \includegraphics[width=0.5\linewidth]{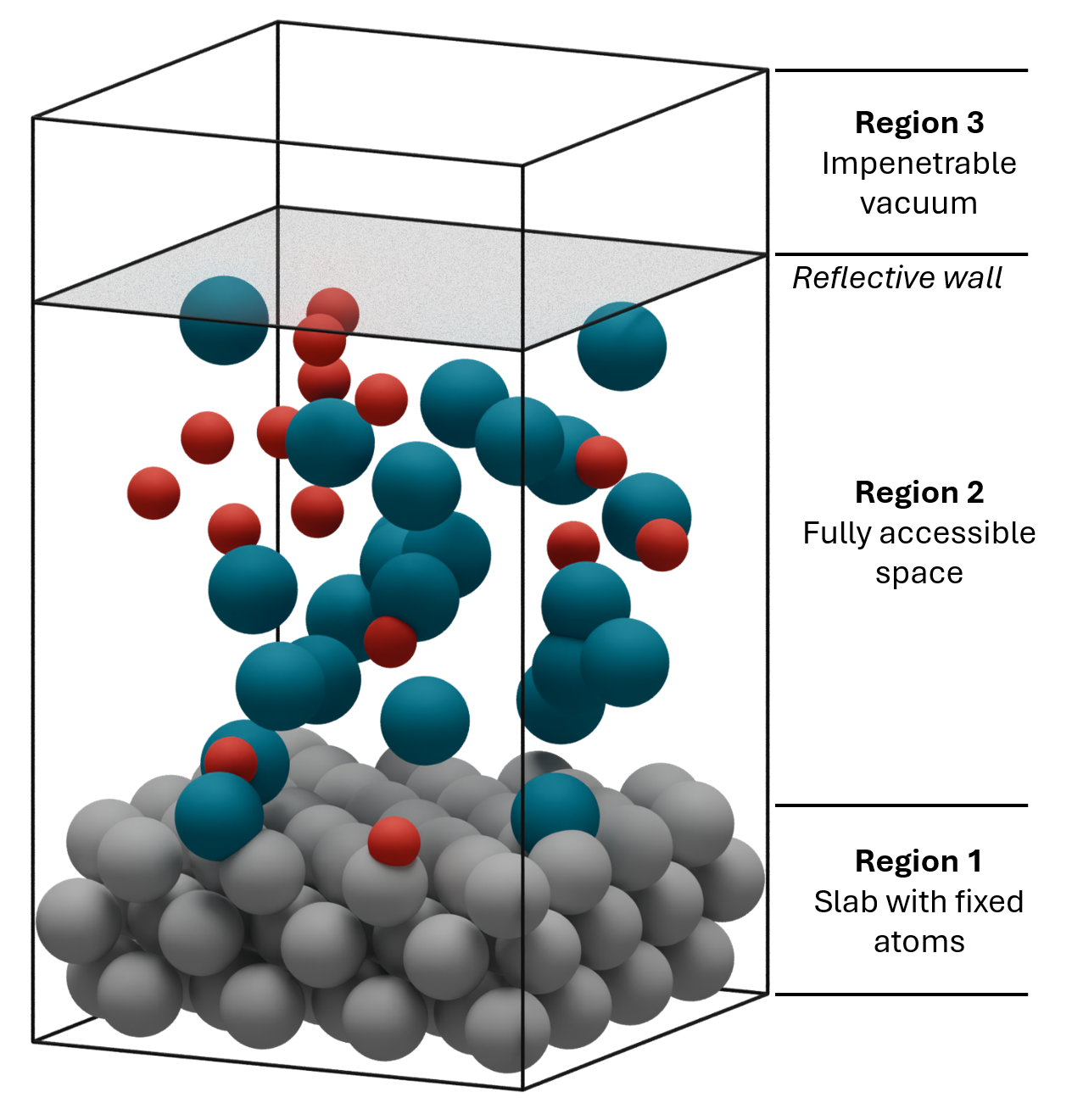}
    \caption{Representative initial configuration of the nested sampling simulation, with relevant simulation cell regions annotated. Region~1 contains the Cu(100) slab with fixed Cu atoms (gray). Region~2 corresponds to the space fully accessible to the mobile Cu (blue) and O (red) atoms. Region~3 is an impenetrable vacuum separated from Region~2 by a reflective wall (gray).}
    \label{fig:main_2}
\end{figure}

The Cu(100) slab model is based on a c$(4\times4)$ ($(2\sqrt{2} \times 2\sqrt{2})\text{R45°}$) surface unit cell, which is large enough to accommodate two missing rows (MRs) along both $x$ and $y$ direction -- sufficient to support a diverse set of reconstruction motifs -- while remaining tractable for nested sampling simulations. Following the approach of Yang \textit{et al.}~\cite{yang2024surface}, the simulation box is divided into three distinct regions, as illustrated in Fig.~\ref{fig:main_2}.

\begin{itemize}
    \item \textbf{Region~1} constitutes the base slab, composed of three Cu bulk-like layers whose atomic positions remain fixed throughout the simulation, thereby providing a rigid substrate that mimics the underlying bulk crystal.
    \item \textbf{Region~2} extends 15~\AA{} between the top of the base slab and a reflective wall. At the start of the simulation, the mobile atoms of each walker $\mathbf{x}_i$ are initialized as an ideal gas by assigning random positions within region 2. This ensures that all walkers are uniformly distributed in configuration space. Unlike the atoms in region~1, those in region~2 are subject to the complete set of Monte Carlo moves.
    \item \textbf{Region~3} consists of an impenetrable vacuum layer whose thickness exceeds the MACE interatomic potential cutoff ($r_{\mathrm{cut}} = 5$\,\AA{}), ensuring that the periodic boundary conditions do not produce spurious interactions between free particles in region~2 and the bottom of the base slab.
\end{itemize}

The number of walkers was set to 60 per mobile atom, yielding a total of 1440 walkers (24 mobile atoms) for pure Cu and 2400 (40 mobile atoms) for Cu-O systems, respectively. Although this is slightly below the 80 walkers per atom used for Lennard-Jones surfaces~\cite{yang2024surface}, we obtain consistent results across independent runs (see Results and Discussion), indicating that this choice offers a suitable compromise between sampling accuracy and computational cost. Furthermore, a simulation is considered converged once the latest sampled configuration has a negligible contribution to the ensemble at 25\,K:
\begin{equation}
P(\mathbf{x}_{\rm new}) < e^{-10} \, \max\limits_{\mathbf{x}} P(\mathbf{x}) \, , \quad  T = 25\,{\rm K} \quad .
\end{equation}
Additional details about the simulation setup are provided in the SI section 2.

\section{Results and Discussion}
\label{sec:results}

By establishing nested sampling in combination with an accurate MACE potential for complex surface reconstructions, new insights into the oxidation of Cu(100) are obtained. As a first step, the MRR is recovered in the presence of oxygen and its absence confirmed without, identifying the reconstruction as an oxygen induced phenomenon. Secondly, the evolution of the MRR with temperature is investigated, resolving the formation and proliferation of defect states which leads to an order-disorder transition.

\subsection{The Oxygen Induced Missing Row Reconstructions}

To evaluate the role of oxygen in forming the MRR, we investigate two systems with compositions chosen to match the experimentally observed MRR Cu content ($\theta_{\rm Cu} = 0.75$~ML) \cite{liu2024modeling}. The two systems, however, differ with respect to the presence of oxygen leading to varying numbers of mobile atoms (24 Cu atoms for pure Cu with $\theta_{\rm Cu} = 0.75$~ML and 24 Cu + 16 O atoms for Cu-O with $\theta_{\rm Cu} = 0.75$~ML, $\theta_{\rm O} = 0.5$~ML). Consequently, the systems also differ in the accessible phase space volume and the number of nested sampling walkers is adjusted accordingly to resolve the phase space accurately (see Section~\nameref{sec:methods}). The simulations yield an ensemble of 0.5 million configurations for the Cu-only system and 1.4 million configurations for Cu-O. To assess statistical robustness, each simulation was repeated four times with independent initial configurations.

\begin{figure}[H]
    \centering
    \includegraphics[width=1.0\linewidth]{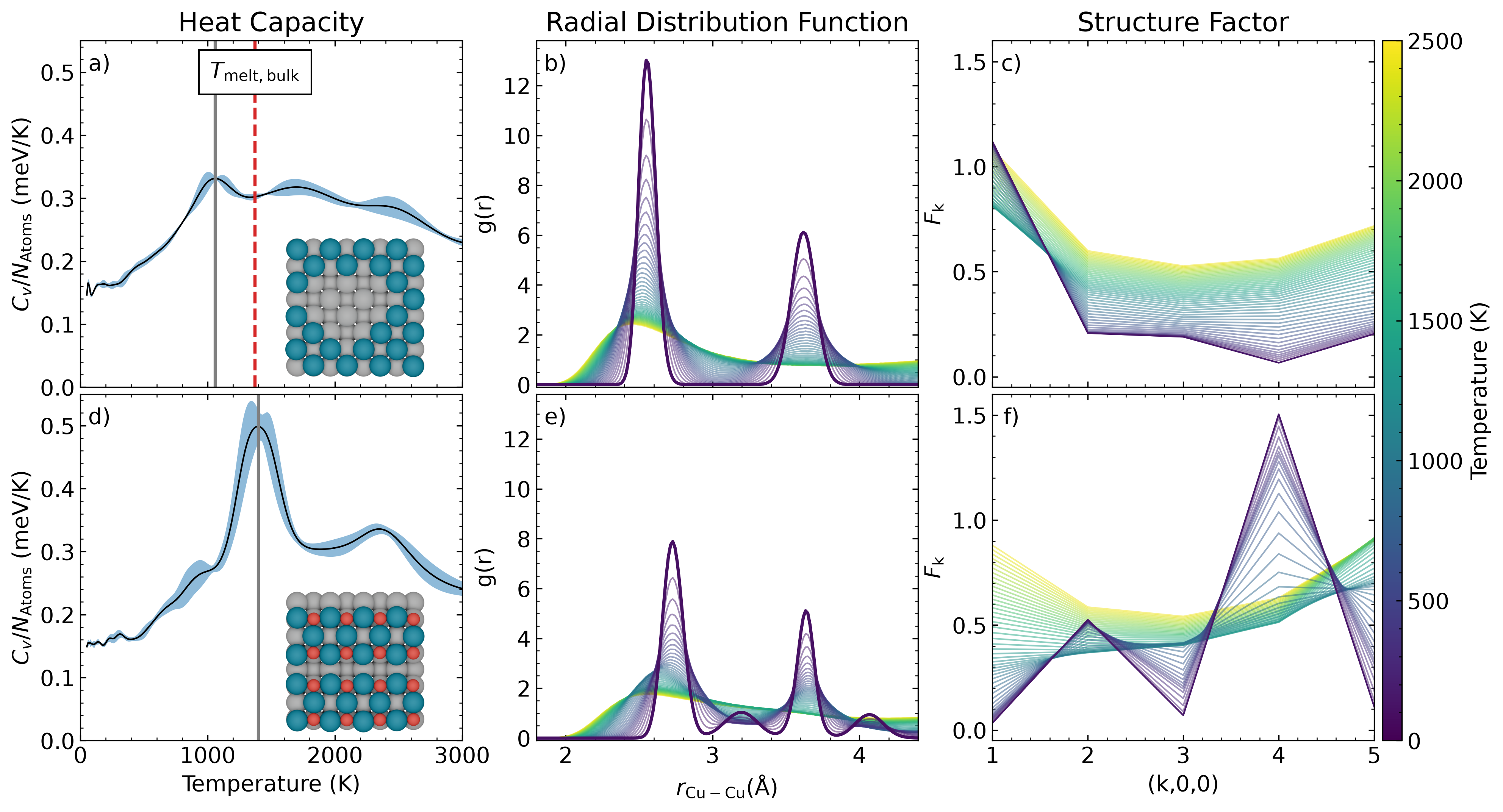}
    \caption{Thermodynamic properties and atomic observables for the $\theta_{\rm Cu} = 0.75$~ML slab (a--c) and the $\theta_{\rm Cu} = 0.75$~ML, $\theta_{\rm O} = 0.5$~ML slab (d--f): heat capacity (a, d), Cu-Cu radial distribution function (b, e), and Cu structure factor (c, f). For all properties, the average over four independent simulations is shown and for the heat capacity additionally the standard deviation (blue shades). The insets in (a, d) show a top view of the global minimum structures, where the mobile Cu atoms are colored blue, fixed Cu in gray and O atoms in red. The red line in (a) shows the MLIP based bulk melting point estimated from the cohesive energy~\cite{guinea1984scaling}. The gray lines in (a, d) mark the surface melting points.}
    \label{fig:main_3}
\end{figure}

\paragraph{Global Minima.}
A characteristic observable of nested sampling simulations are the configurations sampled at the end of the simulation (last 5~\%). These are the configurations with the lowest energy and thus reveal whether the simulation converged to a single minimum energy basin or if several local minima exist with comparable energies. For the investigated systems and convergence criterion (at $T=25$~K), the four independent simulations for both the Cu-only and Cu-O cases yield a single minimum energy basin. The minima differ between the two cases and include translational, rotational, and vibrational variants, respectively. The tight convergence criterion together with the reproducibility between simulation runs justify treating these surface structures as the global minima of their corresponding systems. For the Cu-only system, the global minimum is a heart-shaped anti-island structure (Fig.~\ref{fig:main_3}a) that lacks any MR features. By contrast, the Cu-O simulations result in the experimentally known ideal MRR (Fig.~\ref{fig:main_3}d), demonstrating that nested sampling can reliably recover non-trivial surface reconstructions.

\paragraph{Phase Transitions.}
A key thermodynamic property that can be computed  from the partition function and is, therefore, directly accessible from nested sampling simulations is the heat capacity (see SI section 5). Peaks in the heat capacity correspond to large changes in phase space volume and thus coincide with phase transitions. Beyond its physical significance, the heat capacity can be used to validate the choice of nested sampling parameters. An insufficient number of walkers or inadequate decorrelation lead to noticeable run-to-run shifts in peak positions and distortions of the curve shape~\cite{partay2021nested}. Heat capacity curves are computed for each independent nested sampling run and the corresponding average together with the standard deviation is shown in Figs.~\ref{fig:main_3}a) and~d). The small standard deviation across the four independent simulations indicates that our choice of simulation parameters is appropriate and yields a well-converged heat capacity.

For both systems, the heat capacity exhibits two major peaks: the transition from the gas to a condensed surface layer (above $4000$~K, see SI Fig.~S3), and the surface melting point ($1060$~K for the pure Cu surface and $1400$~K for the CuO surface, gray lines in Figs.~\ref{fig:main_3}a and~d). As common for surfaces\cite{de2023melting}, the melting temperature of the pure Cu surface lies $313$~K below the experimental bulk melting temperature of $1357$~K.~\cite{haynes2016crc}, which is in close agreement with the  bulk melting point $T_{\rm bulk} =1373$~K estimated for the MLIP from the cohesive energy~\cite{guinea1984scaling}. Although less pronounced, premelting is also observed for the Cu-O surface, with a melting point $105$~K below the \ce{Cu2O} bulk melting point ($T_{\rm bulk} =1505$~K, experimental~\cite{haynes2016crc}). Both surface melting peaks exhibit a shoulder on their low-temperature side, which can be assigned to an order--disorder transition between the global minimum structures and metastable defect states as we elaborate later. This shoulder is considerably more pronounced for the Cu-O surface, consistent with the expectation that the oxygen-containing system possesses a richer energy landscape, with a greater number of accessible low-energy local minima than the Cu-only surface. In addition to the major phase transitions, minor, less obvious peaks in the heat capacity are observed at $2320$~K and $1700$~K for the Cu-only surface, and $2350$~K for the Cu-O surface. To better understand the origin of these peaks, the structural features of the surface reconstructions are examined in more detail.

\paragraph{Structural Features.}
The radial distribution function (RDF) and the structure factor are examined to obtain deeper insight into the structural changes underlying the phase transitions and driving the surface reconstructions. Weighted ensemble averages for these atomic observables are evaluated across the entire temperature range from each nested sampling run (see SI section 3) and displayed in Fig.~\ref{fig:main_3}.

In Figs.~\ref{fig:main_3}b) and~e), the weighted Cu--Cu RDFs of the Cu and Cu-O surfaces are shown within a temperature range from 25~K (purple) to 2500~K (yellow). At elevated temperatures ($T > 1500$~K), the RDFs of both systems are similar, reflecting comparable short-range liquid-like order of a melted surface layer. At lower temperatures, however, pronounced differences emerge. In the pure Cu system, the RDF has a nearest-neighbor (NN) peak at $r_\text{Cu-Cu} \approx 2.4$~\AA\ and a second NN peak at $r_\text{Cu-Cu} \approx 3.6$~\AA. In the Cu-O system, the NN peak shifts to slightly larger interatomic distances of $r_\text{Cu-Cu} \approx 2.6$~\AA, a direct consequence of oxygen intercalation between neighboring copper atoms. Additionally, instead of a single second NN peak, three peaks can be found in the Cu-O system. One peak is comparable to the pure Cu second NN distance 
and corresponds to the Cu-Cu distance parallel to the MR with $r_\text{Cu-Cu} \approx 3.6$~\AA{}.
The two additional peaks correspond to distances perpendicular to the MR, a shorter one across the MR with $r_\text{Cu-Cu} \approx 3.2$~\AA{} and a longer one within the Cu-O row at $r_\text{Cu-Cu} \approx 4.1$~\AA{}. These observations confirm previous findings, that oxygen intercalation enlarges the Cu--Cu NN distances and introduces surface strain, with the MRR serving as a mechanism for strain relief.\cite{harrison2006adsorbate,lahtonen2008oxygen} The MRR characteristic features only manifest at $T\lesssim 800$~K, well below the surface melting temperature at $T \approx 1400$~K. The onset of the observed short-range order around $T\approx 800$~K coincides with a shoulder visible in the heat capacity curve which we assign to an order--disorder transition. The strain from oxygen intercalation is compensated by the entropy gain from populating numerous MRR defect states, which are further discussed below.

The structure factor shown in Figs.~\ref{fig:main_3}c) and~f) provides complementary information about the periodicity and thus the long-range order present in the system for temperatures ranging from 25~K (purple) to 2500~K (yellow). At elevated temperatures, the structure factor for both systems appears similar and does not exhibit any prominent peaks, consistent with the absence of long-range order in molten systems. At low temperatures, MRR characteristic peaks emerge for the Cu-O system: The two parallel MRs inside our simulation cell result in $L_{\rm cell}/2$ periodicity perpendicular to the MRs, which causes both a $(2,0,0)$ and a $(4,0,0)$ peak. The $L_{\rm cell}/4$ periodicity along the MRs result in a $(4,0,0)$ peak. The ensemble averaged structure factor contains both vertical and horizontal oriented MRRs resulting in the $(2,0,0)$ peak to be smaller than the $(4,0,0)$ peak. In contrast, the Cu-only surface displays a single peak at $(1,0,0)$, indicating the absence of any periodic features along this orientation and thus the absence of the MRR. Together, the RDF and the structure factor indicate that the MRR is a purely oxygen-induced phenomenon driven by strain relief.

Additional evidence for the main phase transitions and peaks in the heat capacity is provided by the height profile of the surface layer (see mean and maximum $z$-positions of the mobile atoms in SI Fig.~S4).
The origin of the minor peaks at $\approx1700$~K for Cu and $\approx2350$~K for CuO remains less clear. However, the structure factor still changes at these temperatures (see Figs.~\ref{fig:main_3}c and f), which suggests that minor changes in the long-range order might be responsible for these peaks.

\newpage
\subsection{Temperature-Driven Missing Row Reconstruction Evolution}

\begin{figure}[H]
    \centering
    \includegraphics[width=0.5\linewidth]{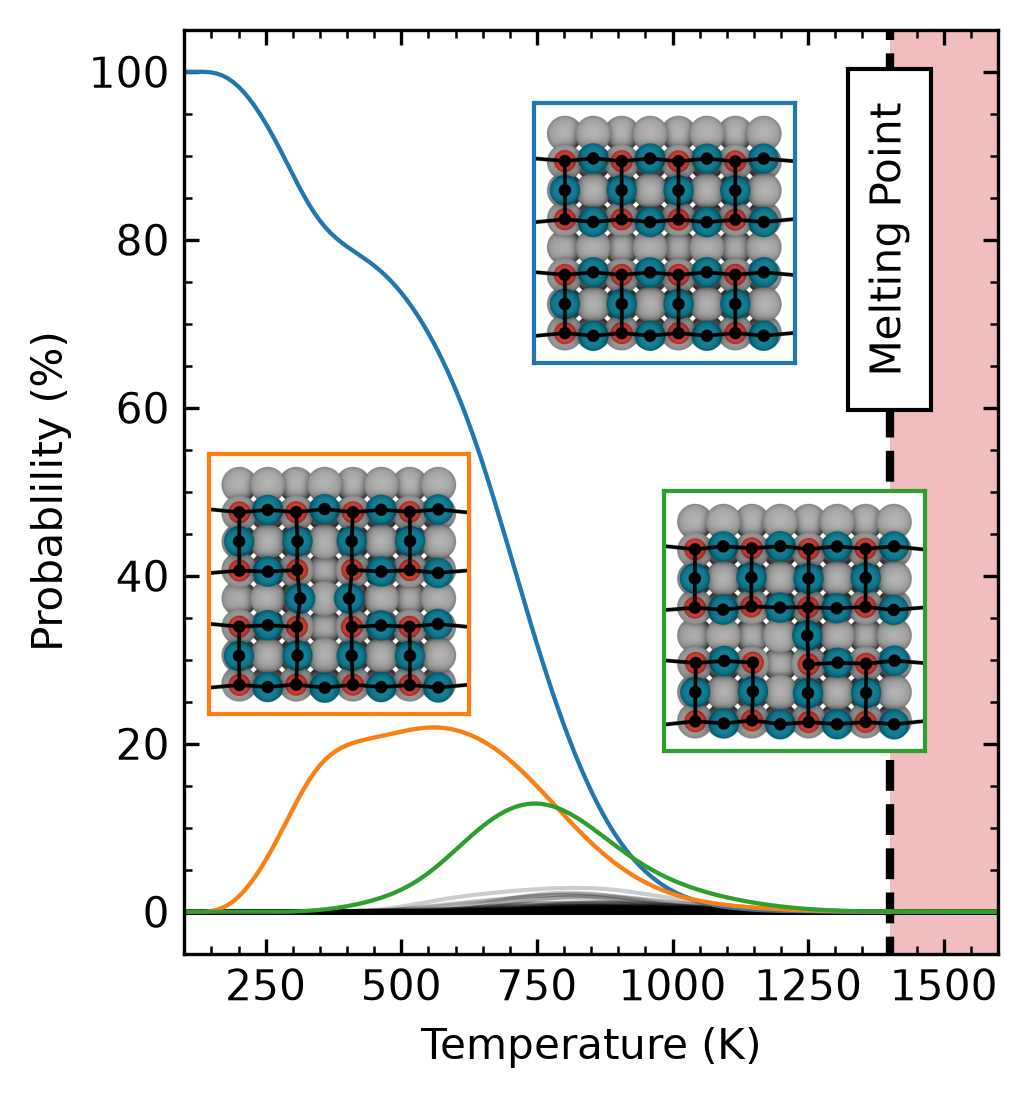}
    \caption{The change in population of the ideal MRR and distinct defect states vs. temperature, averaged over four simulations. The blue line and the corresponding framed structure represent the ideal MRR, while the orange and green lines with their respective framed structures correspond to the two dominant defect states. The probabilities of the remaining structures are colored in gray. While a handful of minor reconstructions shortly reach maximum probabilities $< 5~\%$ at $T\approx 800~{\rm K}$, the probabilities remain predominantly $\ll 0.1\,\%$, appearing as horizontal lines. The different defect structures are identified via canonical graphs, which are visualized as black lines (edges) and points (nodes) in the structures. Mobile Cu atoms are colored blue, fixed Cu atoms in gray and O atoms in red. The dashed black line indicates the surface melting temperature according to the heat capacity and the region above $T_m$ is highlighted in red.}
    \label{fig:main_4}
\end{figure}

The structural evolution of the MRR up to surface melting can be tracked by analyzing the slope and pronounced shoulder in the heat capacity of the Cu-O system in more detail. As noted above, this evolution is likely governed by MRR defects that proliferate into an order-disorder transition. To confirm this assignment and analyze the underlying structures in detail, we employ canonical connectivity graphs as vibrational, rotational, and translational invariant global structural classifiers to capture topologically distinct bond structures of the defect states. In these graphs, Cu and O atoms serve as nodes, and an edge is placed between a Cu--O pair whenever their interatomic distance is less than 2.5\,\AA{}. This threshold lies well above the equilibrium Cu--O bond length of approximately 1.9\,\AA{}, yet remains significantly below the next-nearest Cu--O distance of 3.8\,\AA{}, ensuring that the descriptor is robust against thermal vibrations. The resulting canonical graph representation provides a computationally efficient and reliable means of separating the $\approx5.6$~million sampled configurations from four simulations into over 2.2~million topologically unique bonding patterns. Although solid surface reconstructions often share the same dominating topological graph across multiple configurations, most liquid- and gas-like configurations each correspond to a distinct graph with vanishing individual probabilities. The probabilities of all unique topological graphs are shown in Fig~\ref{fig:main_4} (plots of individual nested sampling runs are presented in SI Fig.~S5). Below the melting temperature, the ideal MRR and two dominant defect structures (blue, orange, and green lines) have a significant contribution while the remaining graph patterns (gray lines) occur only with very low probabilities.

The probabilities of the different topologies in the Cu-O system reveal how the MRR evolves with temperature. Below 300 K, the ideal MRR is the sole dominant configuration (blue line in Fig.~\ref{fig:main_4}), with  two distinct defect states emerging between 300 to 600~K that disrupt the ordered phase. One defect state reaching a probability of 22\% at 600~K is characterized by two Cu atoms bridging neighboring rows, which leads to a partially formed perpendicular MR (orange panel in Fig.~\ref{fig:main_4}). This motif has been previously reported as precursor to domain interconversion \cite{liu2024modeling, baykara2013atom, fujita1996phase}. The other defect state with 8\% probability at 600~K is an intermediate between the MRR and the defect mentioned above, containing a single Cu atom that links neighboring rows (green panel in Fig.~\ref{fig:main_4}). Both defects are already accessible near the MRR onset temperature of 400~K,\cite{lian2017calculations}, their contribution grows with increasing temperature, which would explain why experiments increasingly report MRR defects after annealing at 500 to 600~K, where the predicted MRR population drops to 60\%.\cite{shields2024stm, liu2024modeling, li2022uneven} The predicted defect states resemble the point-like and L-shaped MR patterns reported experimentally, which can be associated with Cu MR mismatch-induced phase boundaries, where orthogonal MRR domains meet. Previously, such boundaries have been attributed to step edges, domain mismatches, or domain interconversion, \cite{liu2024modeling, baykara2013atom} yet our data show that the underlying structural defects are actually thermodynamically competitive at these temperatures. Defect formation and the subsequent surface evolution therefore appear to be thermodynamically driven, in contrast to earlier assumptions of a purely kinetic stabilization. \cite{saidi2012ab} This driving force is consistent with predicted diffusion and nucleation barriers below 1\,eV for the Cu(100)-O surface, which are readily surmounted at annealing temperatures above 400\,K.\cite{lian2017calculations}

The same thermodynamic accessibility extends to the order--disorder transition. It emerges beyond typical annealing temperatures near 800~K, aligning with the shoulder in the heat capacity (see Fig.~\ref{fig:main_3}d) and marking the point where the entropy gain from forming defects outweighs the strain relief from forming ideal MRRs. Above this temperature, the ideal MRR  accounts for $\leq$ 20\% of configurations, while the remaining defect patterns (gray lines in Fig.~\ref{fig:main_4}) each appear with very low probability, so that many states become thermodynamically accessible and collectively define the disordered regime. Notably, the transition occurs at substantially lower temperatures than the onset of surface melting at 1400\,K, implying that the disordered solid phase persists over a broad temperature window of roughly 600\,K. In fact, beyond 1000~K, a high degree of structural disorder is reached, with probabilities of the ideal MRR and both major defect states falling below $5\,\%$ and numerous defect states occurring with an equally low probability, merging to a single black line in Fig~\ref{fig:main_4}. Within this extended disordered regime, we observe a large and diverse pool of distinct surface configurations. In addition, many structures exhibit thermal vibrations of sufficient amplitude to break the connectivity graph, signaling the onset of surface roughening.

\subsection{Relative stability of surface compositions}
\begin{figure}
    \centering
    \includegraphics[width=0.5\linewidth]{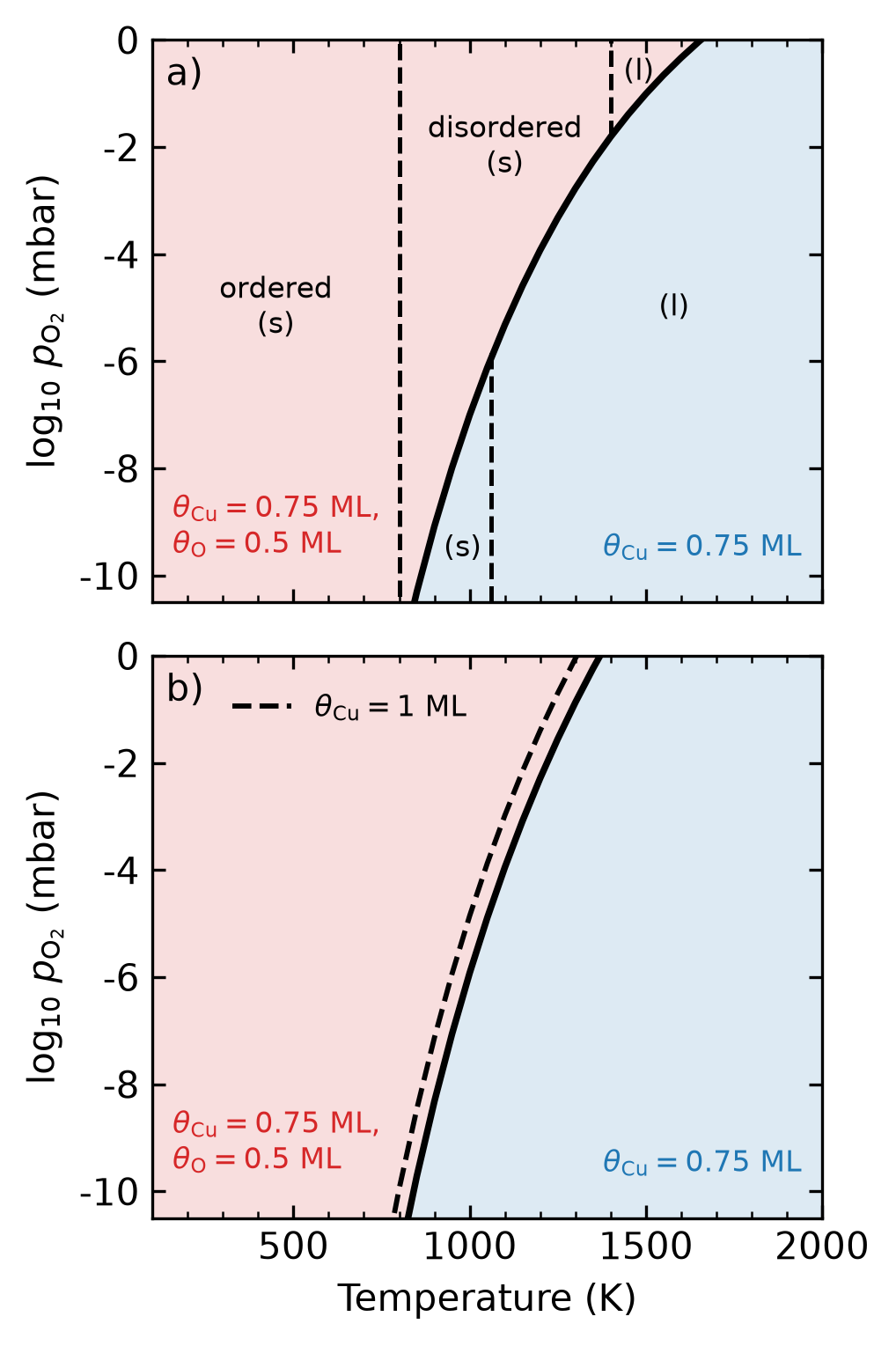}
    \caption{Phase stability of the Cu-only slab ($\theta_{\rm Cu}=0.75$~ML, blue) and the oxidized Cu slab ($\theta_{\rm Cu}=0.75$~ML, $\theta_{\rm O}=0.5$~ML, red) as a function of temperature and oxygen partial pressure. The solid black line marks phase coexistence. (a) Phase diagram based on free energies obtained from nested sampling; the vertical dashed lines indicate the phase transitions of each system within the canonical ensemble (s = solid, l = liquid). (b) Phase diagram based on global-minimum energies, following the conventional \textit{ab initio} thermodynamics approximation. The dashed line labeled $\theta_{\rm Cu}=1$~ML additionally marks the phase boundary between the missing-row reconstructed surface oxide and the clean Cu(100) slab.}
    \label{fig:main_5}
\end{figure}

Another key advantage of nested sampling simulations is that they provide direct access to free energies within the canonical ensemble (see SI sections 5.2). To compare the stability of the two investigated surface compositions, we combine the free energy estimates from nested sampling with a procedure analogous to \textit{ab initio} thermodynamics,\cite{reuter2001composition, rogal2006ab} using a common O$_2$(g) reference. The resulting surface phase diagram is shown in Fig.~\ref{fig:main_5}a), illustrating the relative stability of the two surface compositions as a function of temperature and O$_2$ pressure. The corresponding surface phase diagram obtained with conventional \textit{ab initio} thermodynamics, which is based on $T=0$~K total energies only, is provided in  Fig.~\ref{fig:main_5}b).
The overall trend is comparable and in agreement with a previous study~\cite{saidi2012ab} but the surface phase diagram derived from nested sampling free energies reveals additional insight. First, the transition temperature between the Cu-O and Cu-only systems shows an increase of 30~K to 270~K at low and high O$_2$ pressures, respectively, indicating the stabilization of the Cu-O surface structures through entropy contributions. 
Second, nested sampling provides further information about phase transformations \emph{within} each composition, indicated by the vertical dashed lines in Fig.~\ref{fig:main_5}a). As discussed in the previous sections, the Cu-O system undergoes a transition from the ordered MRR to a disordered state before surface melting. The ordered MRR is stable over a wide $(T,P)$-range and the order-disorder transition occurs before oxygen desorption is favored. It is also noticeable that the solid Cu-only system is only stable at very low O$_2$ pressures while the molten Cu surface layer is dominant at higher temperatures. 

Although the free energy estimates can provide a more complete picture of the surface phase diagram, the analysis is limited to fixed compositions (Cu-O with $\theta_{\rm Cu}=0.75$~ML, $\theta_{\rm O}=0.5$~ML and Cu-only with $\theta_{\rm Cu}=0.75$~ML in this study). In Fig.~\ref{fig:main_5}b), we include the conventional \textit{ab initio} thermodynamics estimate for the defect-free Cu-only surface with $\theta_{\rm Cu}=1.0$~ML, showing a small shift to lower transition temperatures, which is expected as the perfect surface has a lower surface energy. Similarly, in Ref.~\cite{saidi2012ab} an ordered adlayer with $\theta_{\rm Cu}=1.0$~ML, $\theta_{\rm O}=0.25$~ML was found to be stable between the pure Cu surface and the MRR which would slightly reduce the stability range of the ordered and disordered MRR, correspondingly.
To capture the effects of changing composition, the nested sampling approach needs to be extended to the grand-canonical ensemble which goes beyond the scope of our current study and will be explored in the future.

\section{Conclusion}

We have demonstrated that nested sampling combined with a MACE machine-learned interatomic potential reliably recovers complex surface reconstructions without prior knowledge of the relevant phases, transition temperatures, or structural motifs. In contrast to traditional \textit{ab initio} thermodynamic studies, which typically rely on a handful of idealized candidate structures subject to human intuition, nested sampling provides a finite temperature description, including vibrational and configurational entropy effects. Competing minima, defect states, and their thermal occupations emerge naturally from a single simulation. Moreover, unlike most other surface sampling methods, nested sampling grants direct access to the partition function, and thereby to thermodynamic properties such as the heat capacity, atomic observables, and absolute free energy estimates across the entire temperature range, giving access to the temperature-driven thermodynamic surface evolution.

While nested sampling yields a physically accurate description of complex surfaces and their phase space, the approach remains limited by its high computational cost. Although we implemented an extended, highly efficient MC walk entirely on the GPU within the \texttt{torch-sim} framework for a sixfold speedup, only two systems could be simulated with reasonable resources. Both incurred substantial wall times on a single NVIDIA H100, the smaller (24 mobile atoms) requiring roughly one week and the larger (40 mobile atoms) roughly one month. Faster energetics and improved hardware will therefore still be required to extend the method to more demanding applications, such as grand canonical nested sampling simulations for surfaces.

Applied to the early oxidation of the Cu(100) surface, our approach correctly recovers the experimentally observed $(2\sqrt{2}\times\sqrt{2})\,R45^{\circ}$-O MRR as the global minimum of the oxidized surface. Crucially, no MR feature emerges in the absence of oxygen, even at the Cu coverage of the experimentally known MRR, identifying the reconstruction as a purely oxygen-induced phenomenon that relieves the strain introduced by oxygen intercalation between neighboring Cu atoms. Beyond the ground state, our simulations resolve the rich defect chemistry of the MRR in detail, matching defect patterns reported in STM studies of Cu(100) oxidation. We further predict an associated order--disorder transition above typical annealing temperatures at $800$\,K, in which proliferating defects accumulate into a highly disordered surface, with some configurations showing signs of surface roughening. All of these surface transitions are thermodynamically accessible, consistent with the low diffusion and nucleation barriers predicted for the Cu(100)-O surface, and in contrast to hypothesized necessary influences from step edges, domain mismatches, or domain interconversion.

\newpage

\section*{Data Availability Statement}

The gpu-optimized surface nested sampling code is available at \href{https://github.com/Felixrccs/pymatnext/tree/torch_walk}{\texttt{Felixrccs/pymatnext}}~\cite{riccius2026pymatnext} (torch$\_$walk branch), and is based on the \href{https://github.com/libAtoms/pymatnext}{\texttt{pymatnext}} package~\cite{Noam2026pymatnext}. The analysis and plotting scripts are available at \href{https://github.com/Felixrccs/nest_tools.git}{\texttt{Felixrccs/nest$\_$tools}}~\cite{riccius2026analysis}. The simulation data, MACE potential and trainings data is available at \href{https://doi.org/10.5281/zenodo.21703740}{\texttt{Zenodo}}\cite{riccius2026data}.

\section*{Supporting information}

The Supporting Information is available free of charge:

\begin{itemize}
    \item Additional information about MACE training, simulation setup, analysis, walker decorrelation and nested sampling background. Figures showing the simulation setup, decorrelation between lattice sites, z-position and heat capacity. 
\end{itemize}

\section*{Acknowledgements}

The authors thank Livia B. P{\'a}rtay for discussions on Nested Sampling and Chatipat Lorpaiboon for discussions on Galilean Monte Carlo.
FR thanks the ICC at the Flatiron Institute for hospitality while part of this research was carried out there. The computations reported in this paper were in part performed using resources made available by the Flatiron Institute. The Flatiron Institute is a division of the Simons Foundation. Additional computing time was provided by the Max Planck Computing and Data Facility (MPCDF).\\

\noindent
AI tools were used for spell-checking and grammatical corrections.

\printbibliography

\appendix

\setcounter{table}{0}
\setcounter{figure}{0}
\renewcommand{\thefigure}{S\arabic{figure}}
\renewcommand{\theequation}{S\arabic{equation}}
\renewcommand{\thetable}{S\arabic{table}}
\setcounter{equation}{0}
\setcounter{table}{0}
\setcounter{section}{0}
\setcounter{secnumdepth}{3}
\renewcommand\thesection{\arabic{section}}
\renewcommand\thesubsection{\thesection.\arabic{subsection}}

\newcolumntype{Y}{>{\centering\arraybackslash}X}

\newpage

\vspace*{2cm}
\begin{center}
\begin{LARGE}
Supporting Information: \\
Finite-Temperature Thermodynamics of Cu(100) Oxidation: Missing-Row Reconstruction, Defect States, and Order-Disorder Transition from Nested Sampling \\[5ex]
\end{LARGE}
{\Large
Felix Riccius, Karsten Reuter, Hendrik H. Heenen, Jutta Rogal
}
\end{center}

\newpage

\section{MACE training}
We fine-tune our MACE potential starting from a more general Cu-surface-oxide potential\cite{riccius2025out}, adopting the DFT settings of that work for the newly generated plane-wave PBE-based training data\cite{perdew1996generalized, blochl1994projector, kresse1996efficient, kresse1999ultrasoft}. To generate the additional configurations, we employ an active-learning (AL) scheme built around \texttt{EZGA}\cite{lombardi2026ezga}, a global optimization software package. Although the initial MACE potential already performs reasonably well from the first AL generation onward, we add 1200 $4\times4$ Cu(100)-oxide configurations (spanning $\theta_{\rm Cu} \in [0,1]$~ML and $\theta_{\rm O} \in [0,1.5]$~ML) over six AL generations, which yields a small accuracy improvement and, more importantly, enhances robustness that extends to our nested sampling simulation setup. Half of these configurations are drawn from molecular dynamics (MD) simulations and half are local minima, so that the resulting potential accurately describes low-energy configurations while remaining reliable at higher energies. A more detailed description of the AL procedure is available at \href{https://gitlab.mpcdf.mpg.de/fhi-theory/EZGA/-/wikis/Examples/MLIP-Active-Learning}{EZGA/Active-Learning}\cite{jml2026ezga-learning}. The resulting potential achieves a test set root-mean-square error (RMSE) of $1.7\,\mathrm{meV/atom}$ on energies and $45\,\mathrm{meV/\text{\AA}}$ on forces.

\section{Simulation Setup}

\begin{figure}[htb]
    \centering
    \includegraphics[width=1.0\linewidth]{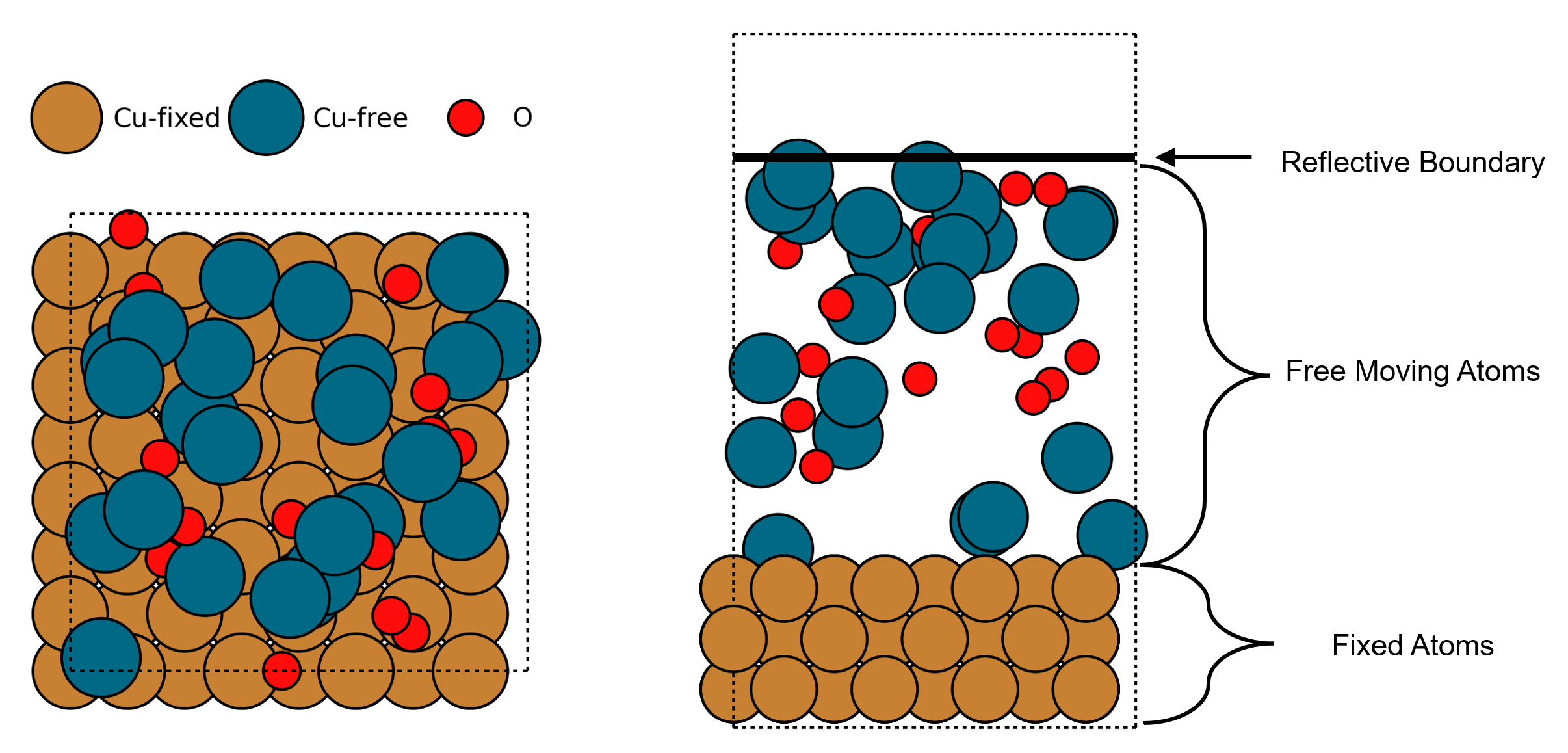}
    \caption{Top and side views of the c$(4\times4)$ simulation cell. The copper-colored Cu atoms remain fixed during the simulation, while the blue-colored Cu atoms and the O atoms (red) are free to move up to the reflective boundary. The setup follows Yang \textit{et al.}\cite{yang2024surface}}
    \label{fig:SI-setup}
\end{figure}
\noindent
We use a c$(4\times4)$ cell, which accommodates two parallel missing rows along both the $x$ and $y$ directions. The three bottom layers of the slab remain fixed throughout the simulation (see Fig~\ref{fig:SI-setup}). To avoid spurious interactions between the moving atoms and the bottom of the fixed slab, we place a reflective wall $15~\text{\AA}$ above the fixed slab, followed by a $5~\text{\AA}$ vacuum region that the atoms cannot enter. To prevent unphysical behavior of the MACE potential at short interatomic distances ($d \to 0~\text{\AA}$), we enforce hard-sphere constraints throughout the simulation: $d_{\rm Cu-Cu} > 1.7~\text{\AA}$, $d_{\rm Cu-O} > 1.3~\text{\AA}$, and $d_{\rm O-O} > 0.9~\text{\AA}$. We use 60 walkers per atom, yielding 1440 walkers for the \ce{Cu24} overlayer and 2400 for the \ce{Cu24O16} overlayer.

\section{Analysis}
The complete analysis and plotting scripts for Figs.~1, 3, and 4 are available at \href{https://github.com/Felixrccs/nest_tools.git}{Nested-Sampling-Analysis}\cite{riccius2026analysis}. The heat capacity is evaluated using the \texttt{pymatnext} package~\cite{Noam2026pymatnext}. The radial distribution function (RDF) and the structure factor are calculated for Cu--Cu pairs only to allow a direct comparison between the \ce{Cu24} and \ce{Cu24O16} simulations. The RDF is computed only within the two dimensions parallel to the surface, since we focus on a temperature range in which our systems form either monolayer-thick films or near-monolayer-thick molten variants thereof.

\section{Walker Decorrelation}
\subsection{Implementation}
To reduce the computational cost and use the GPU more efficiently, we transferred the Monte Carlo (MC) walk to \texttt{PyTorch}, following the \texttt{torch-sim} framework. Instead of using multiple MPI instances, we apply the MC walk to multiple walkers in a batched manner on a single GPU. Our torch-based MC walk proceeds as follows:
\begin{enumerate}
    \item A single MC move is randomly selected for all $N$ walkers.
    \item This allows us to draw a single random number tensor or displacement tensor for all $N$ walkers. The walkers can be propagated in parallel as a batch, avoiding explicit loops.
    \item The move is subsequently accepted or rejected for each walker individually.
\end{enumerate}

\noindent
The GPU-optimized implementation yields an $\approx 6\times$ speedup relative to an MPI-parallelized run using the ASE--MACE calculator, with both benchmarks performed on an NVIDIA H100. This speedup applies to the system sizes considered in this work and is consistent with the efficiency gains reported for the \texttt{torch-sim} framework \cite{cohen2025torchsim}.

\subsection{Galilean Monte Carlo (GMC)}

The main MC move we use is GMC \cite{skilling2012bayesian}. At the start of a GMC move, a displacement vector $\mathbf{d} \in \mathbb{R}^{3N}$ is drawn from a normal distribution, $\mathbf{d} \sim \mathcal{N}(0, \sigma^2)$. The atomic positions $\mathbf{x}$ are then propagated along $\mathbf{d}$ for $M$ steps. Whenever the configuration leaves the confined phase space ($E(\mathbf{x}) > E_{\mathrm{limit}}$), the displacement vector is reflected back towards the confined phase space using the atomic forces $\mathbf{F}(\mathbf{x})$. The move is accepted if all $M$ steps complete without two consecutive reflections, and rejected otherwise --- the latter indicating that the walker is trapped in a narrow region of configuration space. The step size of the GMC move, controlled by the width $\sigma$ of the proposal distribution for the displacement vector, is adjusted throughout the simulation to maintain a target acceptance ratio.

\subsection{MC-moves}
For each nested-sampling iteration, we apply the MC walk to a batch of 12 walkers and stop the walk once its length exceeds 20 steps. We use a diverse set of MC moves, which are randomly selected with a probability proportional to the weight $W$:
\begin{itemize}
    \item \textbf{GMC} \cite{skilling2012bayesian}: Our torch-based GMC implementation follows the \texttt{pymatnext} package. A single GMC move consists of 10 displacement steps and is rejected if the displacement vector has to be reflected twice in a row. The standard deviation of the displacement vector is tuned on the fly to achieve an acceptance ratio between 50\% and 75\%. ($W = 7.5$)
    \item \textbf{Random Displacement}: A randomly chosen atom is assigned a new random position. ($W = 1$)
    \item \textbf{ID Swap}: If more than one species is present, the identities of two randomly chosen atoms of different species are swapped. ($W = 1$)
    \item \textbf{Lattice Site Step}: A random atom is moved in the $x$ and $y$ directions by $X \in \mathcal{Z}^+$ times the surface lattice spacing. ($W = 2$)
    \item \textbf{Lattice Up/Down Step}: A random atom is moved in the $x$ and $y$ directions by $X + 0.5$ times the surface lattice spacing, combined with a displacement of one layer up or down in the $z$ direction. ($W = 2$)
\end{itemize}

\begin{figure}[!htb]
    \centering
    \includegraphics[width=0.9\linewidth]{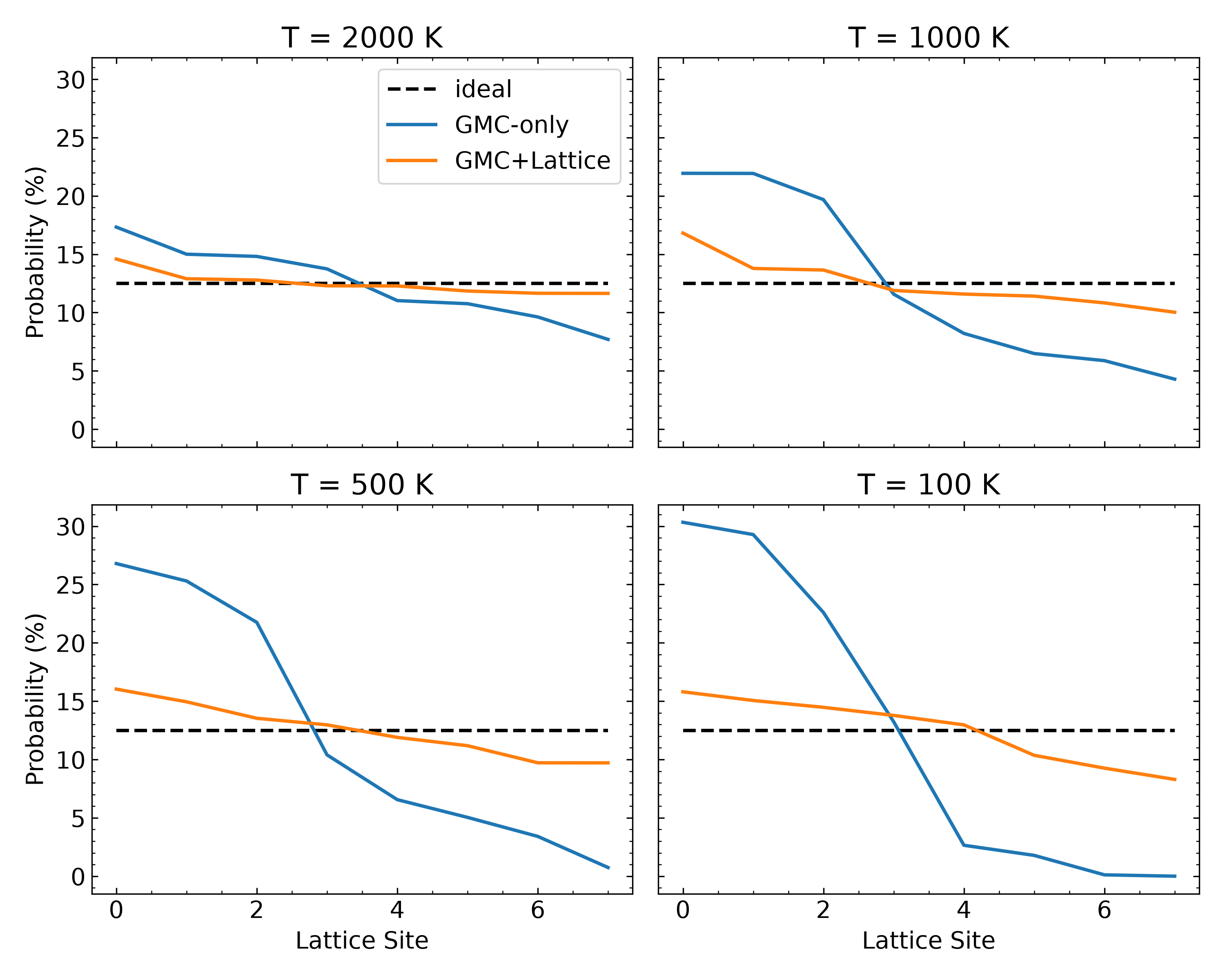}
    \caption{The distribution of a single adsorbate over the 8 degenerate hollow sites on an c$(2\times2)$) Cu(100) surface. The x-axis shows the lattice sites sorted by their probability (y-axis). The dashed black line shows the ideal "equal" distribution while the colored lines show the distribution extracted from two nested sampling simulations with GMC as the only MC move (blue) and GMC in combination with \textbf{Lattice Site Steps}.}
    \label{fig:decorrelation}
\end{figure}

The diverse set of MC moves improves walker decorrelation, whereas GMC alone fails to capture the complex, multimodal potential-energy surface of the slab system used in this work. An example of this is shown in Fig.~\ref{fig:decorrelation}: a single adatom on a c$(2\times2)$ Cu(100) surface should appear in each of the eight degenerate hollow sites with equal probability. A nested sampling simulation using GMC as the only move yields an uneven distribution over the lattice sites. Especially at low temperatures, samples are concentrated on a few lattice sites while others remain unpopulated. The addition of \textbf{Lattice Site Step} moves alleviates this problem by connecting phase-space regions that are separated by a barrier GMC cannot overcome.

Additionally, rarer artifacts we encountered in nested sampling simulations based on GMC only include isolated atoms becoming trapped in vacuum and the formation of adatom-vacancy pairs. These artifacts can be avoided by adding \textbf{Random Displacement} and \textbf{Lattice Up/Down Step} moves, respectively.

\clearpage

\section{Nested Sampling}

The nested sampling background originates from \cite{baldock2017classical,partay2021nested}, the application and implementation is available at \href{https://github.com/Felixrccs/nest_tools.git}{Nested-Sampling-Analysis} \cite{riccius2026analysis}.

\subsection{Partition Function, Heat capacity and Ensemble averages}
The initial configuration-space volume of a fully random initialization is
\begin{equation}
\label{eq:gamma0}
    \Gamma_0 = V^N \quad,
\end{equation}
where $V$ is the volume of the simulation cell and $N$ the number of particles. Because the particles in this work are treated as hard spheres, the accessible $V$ is has to be reduced by the exclusion volume. The exact canonical partition function can then be expressed as
\begin{equation}
\label{eq:Z_full}
    Z(N,V,\beta) = A \int_{0}^{\Gamma_0} d\Gamma \, \exp\!\left(-\beta E(\Gamma)\right) \quad,
\end{equation}
with $\beta = 1/k_\mathrm{B}T$ and
\begin{equation}
    A = \frac{1}{N!h^{3N}} \left( \frac{2\pi m}{\beta} \right)^{\frac{3N}{2}} \quad,
\end{equation}
where $h$ is the Planck constant and $m$ the particle mass. Using nested sampling, we approximate the configurational integral as
\begin{equation}
    \int_{0}^{\Gamma_0} d\Gamma \, \exp\!\left(-\beta E(\Gamma)\right) \approx \sum_i w_i \exp\!\left( -\beta E(\mathbf{x}_i) \right) \quad,
\end{equation}
with
\begin{equation}
    w_i = \Gamma_0 \left( \left[ K/(K+1) \right]^i - \left[ K/(K+1) \right]^{i+1} \right) \quad,
\end{equation}
where $K$ is the number of live points (walkers) and $\mathbf{x}_i$ the configuration removed at iteration $i$. Both $A$ and $\Gamma_0$ enter as common prefactors in canonical simulations and therefore cancel out when calculating the heat capacity or ensemble averages. The heat capacity can be calculated with

\begin{equation}
    C_V(\beta) = k_{\mathrm{B}} \beta^2 \left( \frac{\partial^2 \ln Z}{\partial \beta^2} \right)_{N,V} \;. \label{eq:heat_capacity}
\end{equation}
The probability of observing the sample $i$ at a given temperature can be expressed as
\begin{equation}
    P(\mathbf{x}_i,\beta) = \frac{w_i \exp\!\left( -\beta E(\mathbf{x}_i) \right)}{Z(\beta)} \;, \label{eq:probability}
\end{equation}
which makes ensemble averages of any atomic observable $A$ directly accessible via 
\begin{equation}
\langle A \rangle(\beta) = \sum_i P(\mathbf{x}_i,\beta) \, A(\mathbf{x}_i)\quad.    
\label{eq:ensemble_averages}
\end{equation}
A key advantage of this formulation is that the entire set of nested sampling configurations can be reweighted to arbitrary temperatures, so that the full temperature dependence of atomic observables is captured in a single nested sampling run without additional simulations.

\subsection{Free Energy}

Since all terms in Eq.~\eqref{eq:Z_full} can be evaluated or approximated, an absolute value of $Z(N,V,\beta)$ is accessible. The Helmholtz free energy follows as
\begin{equation}
\label{eq:free}
    F(N,V,T) = -k_\mathrm{B} T \, \ln Z(N,V,\beta) \quad.
\end{equation}
For condensed-phase systems and within the considered pressure range, $F$ can be considered a good approximation to the Gibbs free energy, as the $pV$ term becomes negligible.~\cite{rogal2006ab}

To compare the stability of different surfaces, their respective surface free energy $\gamma$ needs to be computed, defined as
\begin{equation}
\label{eq:surface_g}
\gamma_{\rm Cu/Cu-O} (p,T) = \frac{G_{\rm Cu/Cu-O}^{\rm slab}(p,T) - N_{\rm Cu}\mu_{\rm Cu}^{\rm bulk}(p,T) - N_{\rm O}\mu_{\rm O}^{\rm (g)}(p,T)}{A} -\gamma_{\rm core}(p,T) \quad ,
\end{equation}
where $G_{\rm Cu/Cu-O}^{\rm slab}(p,T)$ is the Gibbs free energy of the pure-Cu and Cu-O slab systems, respectively, $N_{\rm Cu}$ and $N_{\rm O}$ are the number of Cu and O atoms in the two systems, $\mu_{\rm Cu}^{\rm bulk}(p,T)$ and $\mu_{\rm O}^{\rm (g)}(p,T)$ are the chemical potentials of bulk Cu and oxygen gas, $A$ is the surface area, and $\gamma_{\rm core}(p,T)$ is the surface free-energy contribution of the fixed bottom side of the slab.
The two studied systems contain the same number of Cu atoms and have the same surface area as well as the same $\gamma_{\rm core}(p,T)$ contribution, such that the difference in surface free energy $\Delta \gamma = \gamma_{\rm Cu-O} - \gamma_{\rm Cu}$, indicating which phase is more stable, is proportional to the absolute Gibbs free energy difference
\begin{align}
\label{eq:free_energy}
    \Delta G(p,T) 
    &= G_\mathrm{Cu-O}^{\rm slab}(p,T) - G_\mathrm{Cu}^{\rm slab}(p,T) - N_\mathrm{O}\,\mu_\mathrm{O}^{\rm (g)}(p,T) \\ \nonumber
    & \approx F_\mathrm{Cu-O}^{\rm slab}(T) - F_\mathrm{Cu}^{\rm slab}(T) - N_\mathrm{O}\,\mu_\mathrm{O}^{\rm (g)}(p,T) \quad , 
\end{align}
where $F_\mathrm{Cu-O}^{\rm slab}(T)$ and $F_\mathrm{Cu}^{\rm slab}(T)$ are directly obtained from the nested sampling simulations via Eq.~\eqref{eq:free}.
The oxygen chemical potential is calculated following Refs.~\cite{rogal2006ab, riccius2025out}:
\begin{equation}
    \label{eq:mu_O}
    \mu_{\rm O}^{\rm (g)}(p,T) = \frac{1}{2} E_{\rm DFT, O_2}^{\rm (g)} + \frac{1}{2} E_{\rm ZPE, O_2}^{\rm (g)} + \Delta \mu_{\rm O}^{\rm (g)}(p^\ominus,T) + \frac{1}{2}k_{\rm B}T\ln\left(\frac{p}{p^\ominus}\right) \quad,
\end{equation}
where $E_{\rm DFT, O_2}^{\rm (g)}$ is the DFT energy of an isolated \ce{O2} molecule in vacuum, $E_{\rm ZPE, O_2}^{\rm (g)}$ its zero-point energy, and $p^\ominus = 1~{\rm atm}$ the standard pressure. The remaining $\Delta \mu_{\rm O}^{\rm (g)}(p^\ominus,T)$ term is obtained from thermochemical tables~\cite{chase1998nist}. We deliberately use the DFT energy rather than the MACE energy, because our MACE potential is not specifically trained for \ce{O2} molecules.

\section{\textit{Ab initio} thermodynamics}
Analogous to Eq.~\eqref{eq:surface_g}, the surface free energy in conventional \textit{ab initio} thermodynamics is approximated by using $T=0$~K total energies for the slab and bulk free energies
\begin{equation}
\label{eq:gamma_aitd}
\gamma_{\rm Cu/Cu-O} (p,T) \approx \frac{E_{\rm Cu/Cu-O}^{\rm slab} - N_{\rm Cu}E_{\rm Cu}^{\rm bulk} - N_{\rm O}\mu_{\rm O}^{\rm (g)}(p,T)}{A} -\gamma_{\rm core}(p,T) \quad ,
\end{equation}
where $E_{\rm Cu/Cu-O}^{\rm slab}$ is the total energy of the respective pure-Cu and Cu-O slab, and $E_{\rm Cu}^{\rm bulk}$ is the total energy per atom of bulk Cu. The stability range of each surface is again determined by their difference in surface free energy $\Delta \gamma = \gamma_{\rm Cu-O} - \gamma_{\rm Cu}$ or $\Delta \gamma = \gamma_{{\rm Cu,}\theta=0.75} - \gamma_{{\rm Cu,}\theta=1.0}$, respectively, where $\gamma_{\rm core}(p,T)$ is the same for all systems.
All total energies for the surface slabs and Cu bulk were calculated using the MACE potential.
The oxygen chemical potential was computed according to Eq.~\eqref{eq:mu_O}.

\section{Additional figures}

\begin{figure}[H]
    \centering
    \includegraphics[width=0.9\linewidth]{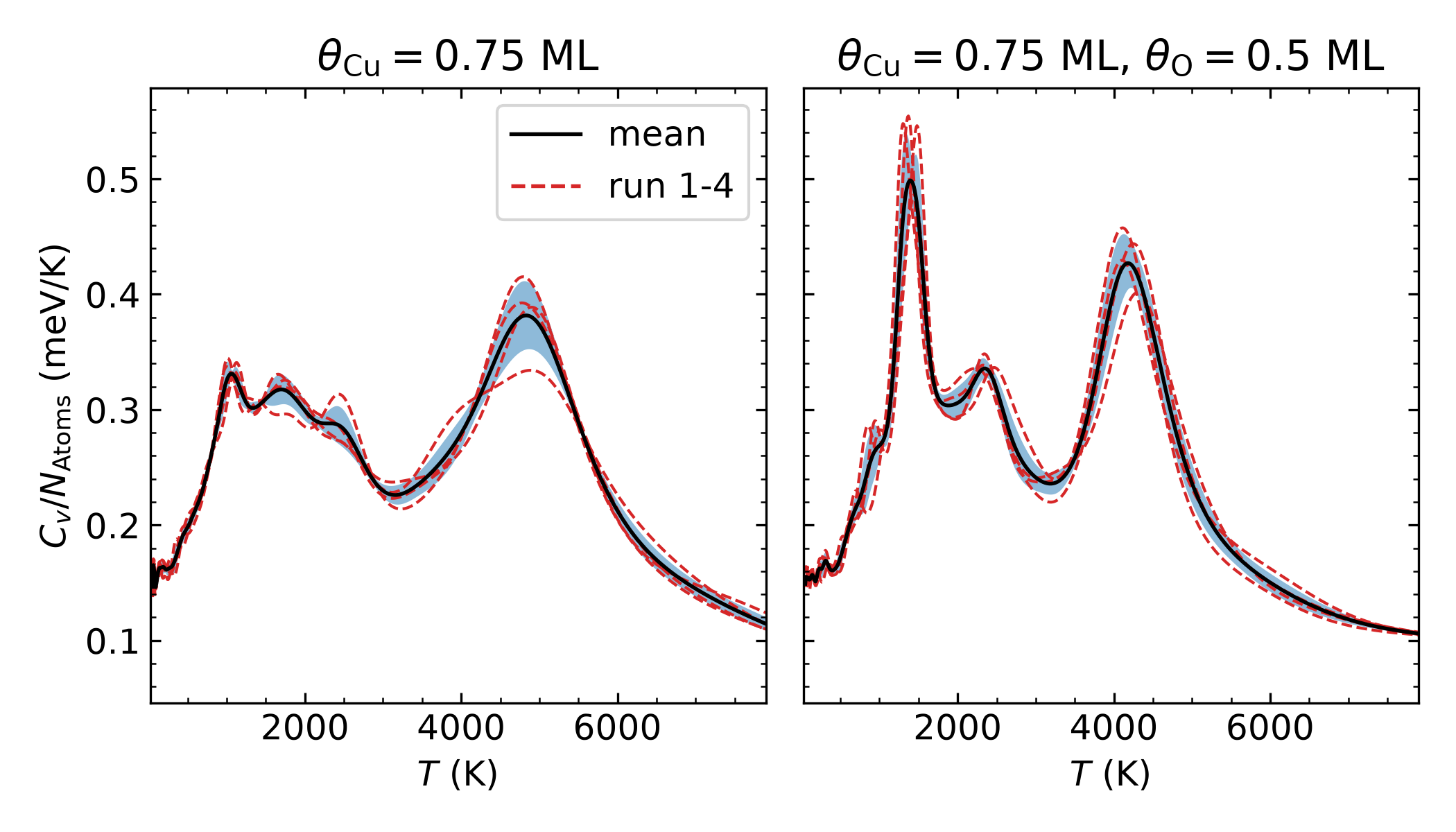}
    \caption{Heat capacity versus temperature for the individual runs (red lines) and their mean (black line) with standard deviation (blue).}
    \label{fig:SI-Cv}
\end{figure}

\begin{figure}[H]
    \centering
    \includegraphics[width=0.9\linewidth]{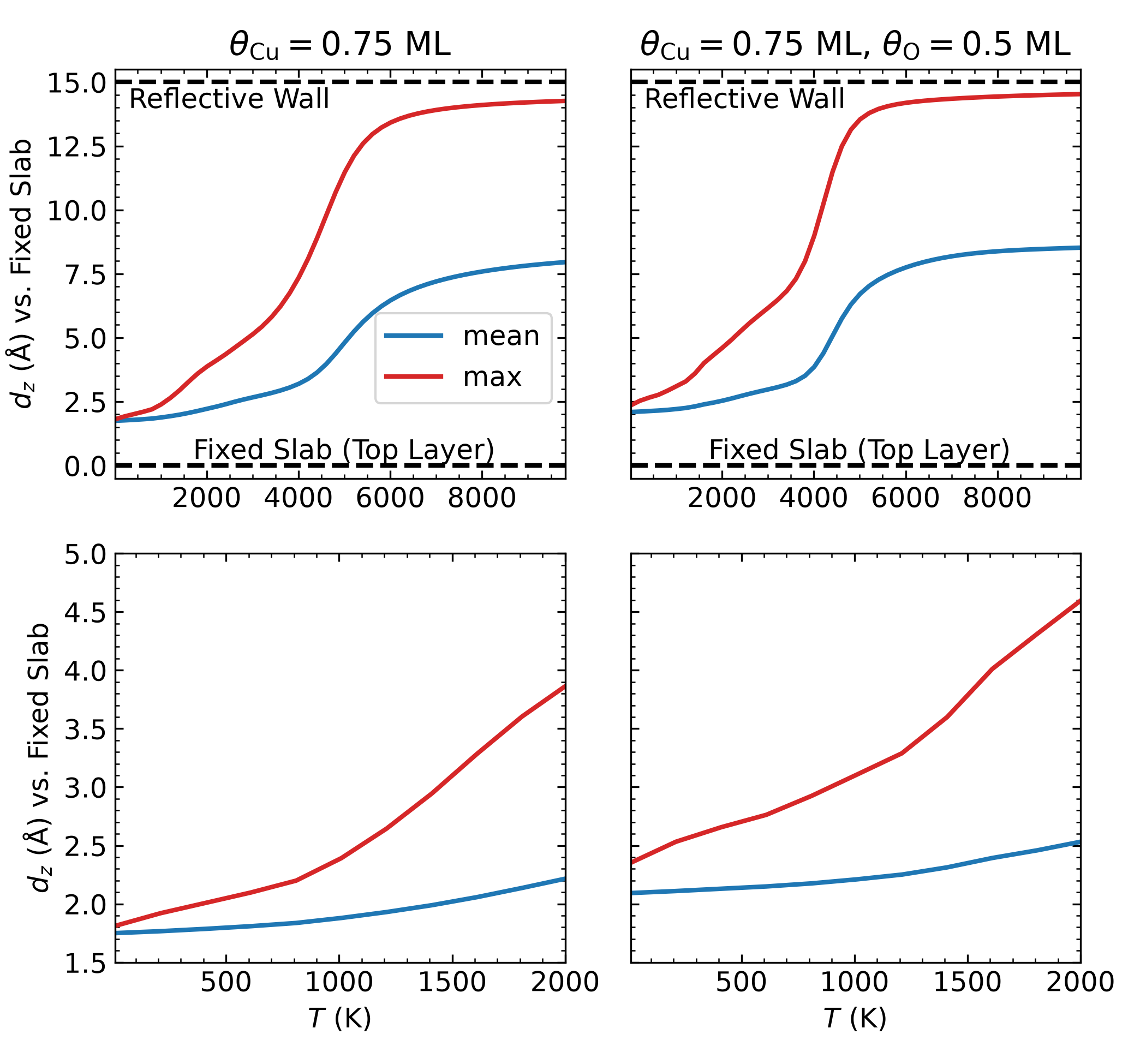}
    \caption{The expectation value of the mean and maximum z-position of the mobile atoms relative to the top layer of the fixed slab versus temperature. The solid, liquid, and gas phases are distinguishable by their different slopes. The lower panels are a magnified view (smaller T-window) to emphasize the different solid/liquid slopes.}
    \label{fig:SI-z}
\end{figure}

\begin{figure}[H]
    \centering
    \includegraphics[width=0.9\linewidth]{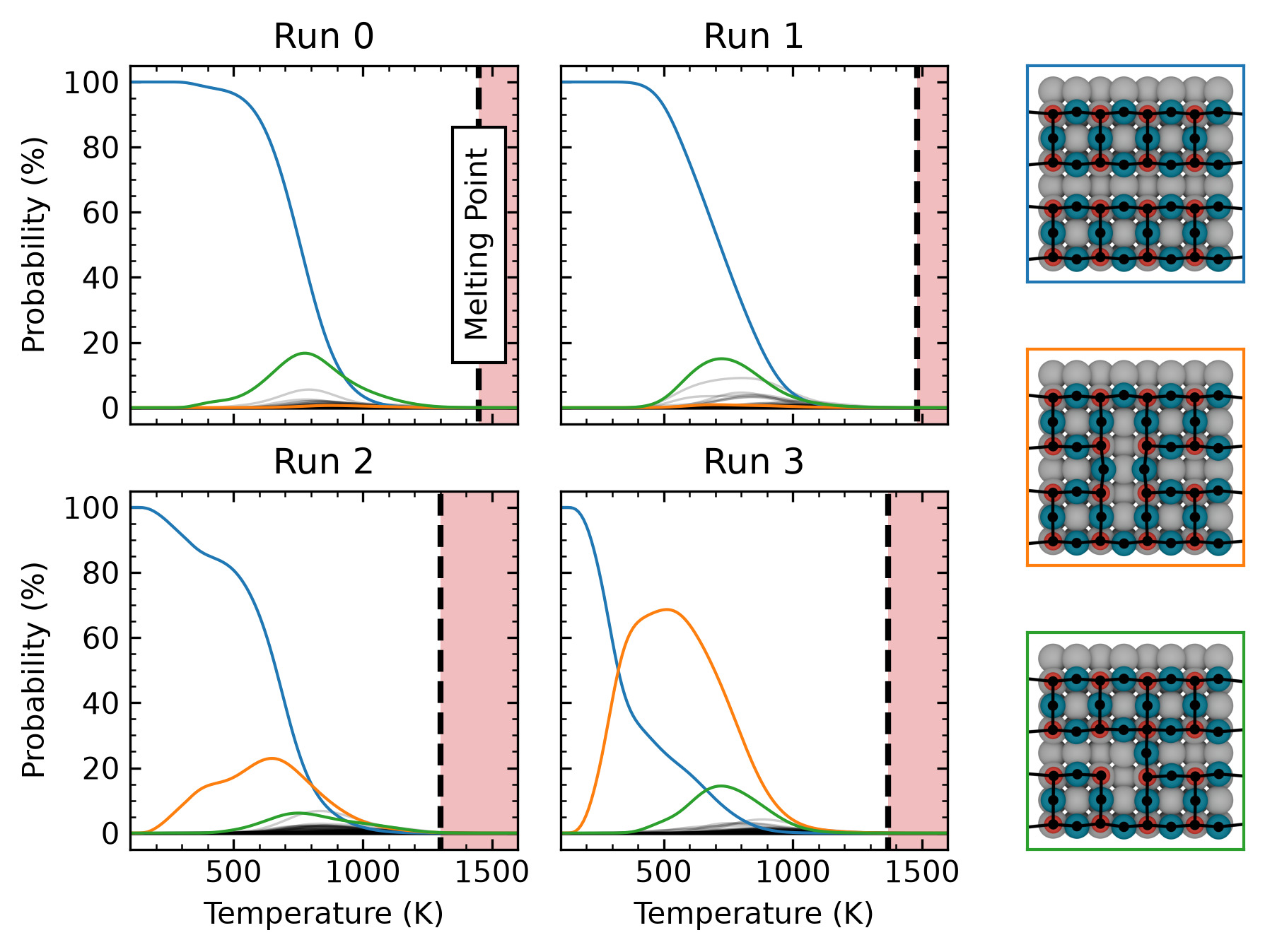}
    \caption{Population of bonding pattern/defect states versus temperature for each individual run. Since nested sampling is a statistical approach, observables can vary between runs, which can be seen in both the melting temperature and the defect population. This variability can be partially offset by averaging over multiple runs, increasing the number of walkers, or reducing walker correlation. Runs 1 and 2 show a phenomenon typical of nested sampling called "extinction", where no walker remains in a secondary minimum basin (here, orange). This illustrates the resolution limits associated with the sampling parameters used in this work.}
    \label{fig:SI-Defects}
\end{figure}

\end{document}